\documentclass[%
 reprint,
 amsmath,amssymb,
 aps,nofootinbib
]{revtex4-2}

\usepackage{graphicx}% Include figure files
\usepackage{dcolumn}% Align table columns on decimal point
\usepackage{bm}% bold math
\usepackage{float}
\usepackage{xcolor}
\usepackage{hyperref}
\begin{document}

\preprint{APS/123-QED}

\title{Experimental verification of arcsine laws in mesoscopic non-equilibrium systems}

\author{Raunak Dey}
\thanks{These two authors contributed equally}
\affiliation{Department of Physical Sciences, IISER Kolkata, India}
\affiliation{School of Physics, Georgia Institute of Technology, USA}
\author{Avijit Kundu}\thanks{These two authors contributed equally}
\author{Biswajit Das}
\author{Ayan Banerjee}\email{ayan@iiserkol.ac.in}
\affiliation{Department of Physical Sciences, IISER Kolkata, India}%

\begin{abstract}
A large number of processes in the mesoscopic world occur out of equilibrium, where the time course of a system evolution becomes immensely important since it is driven principally by dissipative effects. Non-equilibrium steady states (NESS) represent a crucial category in such systems, where relaxation timescales are comparable to the operational timescales. In this study, we employ a model NESS stochastic system which comprises of a colloidal microparticle, optically trapped in a viscous fluid, externally driven by a temporally correlated noise, and show that time-integrated observables such as the entropic current, the work done on the system or the work dissipated by it, follow the three L\'evy arcsine laws \cite{barato2018arcsine}, in the large time limit. We discover that cumulative distributions converge faster to arcsine distributions when it is near equilibrium and the rate of entropy production is small, because in that case the entropic current has weaker temporal autocorrelation. We study this phenomenon changing the strength of the added noise or by perturbing our system with a flow field produced by a microbubble at close proximity to the trapped particle. We confirm our experimental findings with theoretical simulations of the systems. Our work provides an interesting insight into the NESS statistics of the meso-regime, where stochastic fluctuations play a pivotal role.
\end{abstract}

\maketitle
L\'evy arcsine laws is a set of laws which state that some variables related to the Wiener process - a particular type of stochastic process - have a cumulative arcsine probability distribution. \cite{levy1940certains}. To provide an example of Wiener process, we consider the simplest case - i.e. standard Brownian motion of a microparticle. The particle's position coordinate starts from zero. It has an equally probable incremental step in either direction, which does not depend on its past trajectory -- a key signature of a Markovian process. Also, the increments have a zero-centered Gaussian nature where the variance of the distribution increases linearly with time. For such a process, three variables can be defined: a) the proportion of time the fluctuating variable stays above zero: $T_+=\frac{1}{t}\int_0^t\theta(W(t')-\langle W(t')\rangle)dt'$, where $\theta(x)$ is the Heaviside function which takes values '1' and '0' when its argument is positive and negative, respectively (first arcsine law), b) the ratio of the last time when the variable had changed its sign to the total time:  $T_{last}=sup\{t\in[0,1]:W(t)=0\}$ (second arcsine law) and c) the ratio of time when the variable has attained its global maximum to the total time: $T_{max}$ defined as the variable that satisfies,  $W(T_{max})=sup\{W(t),t\in[0,1]\}$ (third arcsine law). These variables are scaled from $0$ to $1$ ($T_+,T_{last},T_{max}\in [0,1]$), and they can be shown to have a normalized probability density function, $\text{PDF}(T_+/T_{last}/T_{max})=\frac{1}{\pi\sqrt{T(1-T)}}$, as long as the root variable follows a Wiener process. It follows that the cumulative distribution function (CDF) of these three variables follow the arcsine law given by:
\begin{equation}
   \text{CDF}(T)=\int_0^T \text{PDF}(T')dT'=\dfrac{2}{\pi}\text{arcsin}(\sqrt{T}).
\end{equation}
Indeed, arcsine laws are observed over a vast range of stochastic phenomena, which include the fluctuation of stock prices \cite{dale1980arc}, waiting time distributions of human dynamics \cite{baek2008testing}, and leads in competitive sports such as football or basketball \cite{clauset2015safe}.  
 
 Wiener processes also appear in the statistical description of microscopic systems as a coarse-grained description of the effect of the environment on the dynamics of the system. Naturally, arcsine-laws have also been observed in this context. Notable examples include the quantum state of a dressed photon in a fiber probe \cite{saigo2018quantum}, the electron current in cold quantum dots \cite{sanchez2013correlations}, the net number of steps of a molecular motor \cite{schmiedl2008efficiency}, the position fluctuation of a particle in a periodic potential \cite{pigolotti2017generic} and various random walk models \cite{bel2006random}. Recently, there have also been generalizations of arcsine laws to include the behavior of fractional Brownian motion \cite{sadhu2018generalized}, non-Markovian processes \cite{singh2019generalised}, and processes that display anomalous diffusion \cite{akimoto2020aging}.
 
A significant addition to this list of scenarios obeying an arcsine law  was made by Barato {\it et. al.,} in \cite{barato2018arcsine}. They considered Markovian non-equilibrium systems in a stationary state and demonstrated that in the long time, the cumulative distribution of the fraction of time a fluctuating current spends above its average value will follow the first arcsine law. In explicit examples of experimentally and numerically realized systems, it was also shown that the second and the third arcsine laws hold as well. Ref \cite{barato2018arcsine} provides a rigorous proof for the first arcsine law for current fluctuations in non-equilibrium steady states, but the second and third arcsine laws remain a conjecture to date.

The examples considered in \cite{barato2018arcsine}, such as the experimental realization of the Brownian Carnot Engine \cite{martinez2016brownian}, are among well studied models in the context of non-equilibrium thermodynamics of microscopic systems \cite{seifert2012stochastic}. Research in this area (also referred to as \textit{Stochastic thermodynamics}) is largely motivated by the desire to uncover general principles that govern the dynamics and thermodynamics in far-from-equilibrium regimes. A large class of systems studied within this framework, for e.g., electric current flowing across a resistor and heat flux across two thermal sources at dissimilar temperatures in contact,\cite{bustamante2005nonequilibrium}, chemical kinetics inside cells \cite{grima2010effective}, and molecular motors \cite{foglino2019non} operating in non-equilibrium steady states. 
%, in systems such as, Stirling cycles \cite{blickle2012realization}, Brownian ratchets and motors \cite{reimann2002brownian}, periodic motion of optical traps. \cite{mestres2014realization}} 
%Notable results include the Fluctuation theorems \cite{jarzynski2011equalities} and the thermodynamic uncertainty relations \cite{horowitz2020thermodynamic}, which have both been tested in a large number of experimental settings, and their applications range from recovering equilibrium information from non-equilibrium measurements to inferring entropy production from measuring fluctuations [\textcolor{red}{Add references}].
 Yet, it is surprising that no other experimental systems, apart from the ones already looked at in \cite{barato2018arcsine} were studied to test the validity of these arcsine laws. Consequently very little is known about their implications in the non-equilibrium steady state dynamics of microscopic systems. 

In this work, we investigate whether L\'evy arcsine laws are obeyed by empirical currents measured for a time duration $\tau$, in the non-equilibrium steady state of a stochastically driven colloidal system. We further consider the convergence to an arcsine law under manifestly different non-equilibrium settings, characterized in terms of the steady state entropy production rate. Our results show that, in all cases, the convergence to the arcsine law is only asymptotic in $\tau$. We characterize the convergence in terms of a distance function in the space of cumulative distributions, and find that the convergence is \textit{faster} for \textit{near} equilibrium systems with a comparatively small entropy production rate as opposed to  far-from-equilibrium systems. To create different non-equilibrium landscapes we drive our model colloidal probe by external noise, change the driving amplitude of the noise, or change the ambient flow field of the fluid where the probe is embedded by trapping it in the vicinity of a microbubble, and study the change in convergence rates. All findings are validated using trajectory data from experiments and numerical simulations.

\subsection{Model and simulation} The model we consider in this work is a colloidal particle trapped in an optical trap, which has its mean position modulated using the Ornstein-Uhlenbeck process. This model was first experimentally tested in ~\cite{gomez:ssw}, and was extensively studied theoretically from a Stochastic thermodynamic point of view  \cite{Pal:2013wfb, verley2014work, Manikandan:2017awd, manikandan2018exact}. In an earlier work, we used this setup to demonstrate the effectiveness of the TUR based inference scheme for the entropy production rate in the stationary state \cite{manikandan2021quantitative}. Different non-equilibrium conditions were created in this setup by controlling the amplitude of the external driving, and by generating microscopic flows in the background by means of a microscopic air-bubble created in the fluid surrounding the probe using a separate laser beam. The latter is particularly interesting since it is a problem that may mimic real life situations involving micro-flows (for eg., bacteria swimming in microflows \cite{secchi2020effect}). However, it is extremely challenging to obtain the exact nature of the flow field analytically, and thereby model the exact force environment that the probe encounters. On the other hand, the TUR method - by construction - overcomes this problem and lays down an upper bound to the maximum entropy that may be generated, as we demonstrated in Ref.~\cite{manikandan2021quantitative}. In this paper, we proceed to calculate the stochastic currents in a similar manner, and investigate the validity of the arcsine laws in this system.

We first consider the case, when the non-equilibrium conditions are created using only the external OU modulation, which we denote using the variable $\lambda(t)$. In this case, the dynamics of our system with position variable $x(t)$ can be described using a system of overdamped Langevin equations as:
\begin{align}
\label{eq:lgv}
\begin{split}
    \dot{x}(t) &= -\dfrac{x(t)-\lambda(t)}{\tau}+\sqrt{2D}\zeta(t)\\
    \dot{\lambda}(t) &= -\dfrac{\lambda(t)}{\tau_0}+\sqrt{2A}\xi(t)
    \end{split}
\end{align}
In Eq.\ \eqref{eq:lgv}, $D$ is the room temperature diffusion constant and  $\tau=\gamma/k$ is the relaxation time of the harmonic trap, where $k$ is the trap stiffness and $\gamma$ is the drag coefficient related by the Stokes-Einstein relation as $D\gamma=k_BT$. Similarly, $\tau_0$ is the relaxation time of the OU process and $A$ corresponds to its strength. The noise terms $\zeta(t)$ and $\xi(t)$ are Gaussian white noises obeying $\langle\xi(t)\rangle = 0$, $\langle\zeta(t)\rangle = 0$,  $\langle\xi(t)\xi(t')\rangle=\delta(t-t^\prime)$, $\langle\zeta(t)\zeta(t')\rangle=\delta(t-t^\prime)$ and $\langle\xi(t)\zeta(t')\rangle=0$. The OU noise has an exponentially decaying correlation $\langle\lambda(t)\lambda(t')\rangle=A\tau_0\exp(-|t-t'|/\tau_o)$ and a power spectrum that falls off as $f^{-\alpha}$, $\alpha\approx2$. In presence of the microbubble the system can be modelled with a suitable addition of the flow field $(u_d)$, $\dot{x}\rightarrow\dot{x}-u_d$ and by values of $\tau \rightarrow \tau_d$ and $D\rightarrow D_d$ that get scaled as a function of the spatial distance $d$ from the microbubble. (see Ref.~\cite{manikandan2021quantitative}). 

An arbitrary, fluctuating time-integrated current $J_d$ in the steady state of this system can be defined as, 
\begin{align}
\label{eq:3}
    J_d^\tau =\int_{\textbf{x}(0)}^{\textbf{x}(\tau)} \textbf{d}(\textbf{x})\circ d\textbf{x}.
\end{align}
Here $\textbf{x}$ denotes the column vector $[x,\lambda]^T$ and $'\circ'$ denotes the Stratanovich product defined as 
\scalebox{0.8}{$\int_{\textbf{x}(0)}^{\textbf{x}(\tau)} \textbf{d}(\textbf{x})\circ d\textbf{x}= \sum_{i=1}^{\frac{\tau}{\Delta t}-1} \textbf{d}\left(\frac{\textbf{x}_{i\Delta t}+\textbf{x}_{(i+1)\Delta t}}{2} \right)\cdot(\textbf{x}_{(i+1)\Delta t})-\textbf{x}_{i\Delta t}))$}. A particular choice of $\textbf{d}(\textbf{x})$  is  referred to as the thermodynamic force field $\textbf{F}(x)$, given by, 
\begin{align}
\bm{F}(\textbf{x})
= 
\left(\begin{array}{c}
         \frac{\delta  \left(\delta ^2 \theta  (\lambda-x)+\delta  \lambda+\lambda\right)}{D\tau_0 \left(\delta ^2 (\theta +1)+2 \delta +1\right)}  \\
          -\frac{\delta ^2 (\delta  x+x-\delta  \lambda)}{D\tau_0 \left(\delta ^2 (\theta +1)+2 \delta +1\right)}
    \end{array}\right),
\end{align}
where we have introduced the dimensionless variables $\delta=\frac{\tau_0}{\tau}$ and $\theta =\frac{A}{D}$. The corresponding time integrated current is the steady state entropy production $\Delta S_{tot}$. The corresponding entropy production rate ($\sigma$) for this system $\sigma = \frac{\langle\Delta S_{tot}\rangle}{\tau}$ is given by  \cite{Pal:2013wfb,verley2014work,manikandan2018exact}, 
\begin{align}
\label{eq:sigma}
    \sigma = \frac{\delta^2 \theta}{(\delta+1)\tau_0}.
\end{align}
The entropy production rate, measured in units of $k_B s^{-1}$, physically corresponds to the heat dissipated from the system to the environment in the steady state. Up to a factor of the bath temperature, and certain boundary terms, it also equals the cumulative work done on the system ($W$) and work dissipated by the system ($W_d$) given by \cite{manikandan2018exact,verley2014work},
\begin{align}
\label{eq:6}
\begin{split}
    W &=\int_0^\tau -k[x(t)-\lambda(t)]* d\lambda(t)/dt\\
    W_d &=\int_0^\tau \lambda(t)* dx(t)/dt + \text{ Boundary terms}
    \end{split}
\end{align}

\begin{figure*}[t]
\includegraphics[width=18cm]{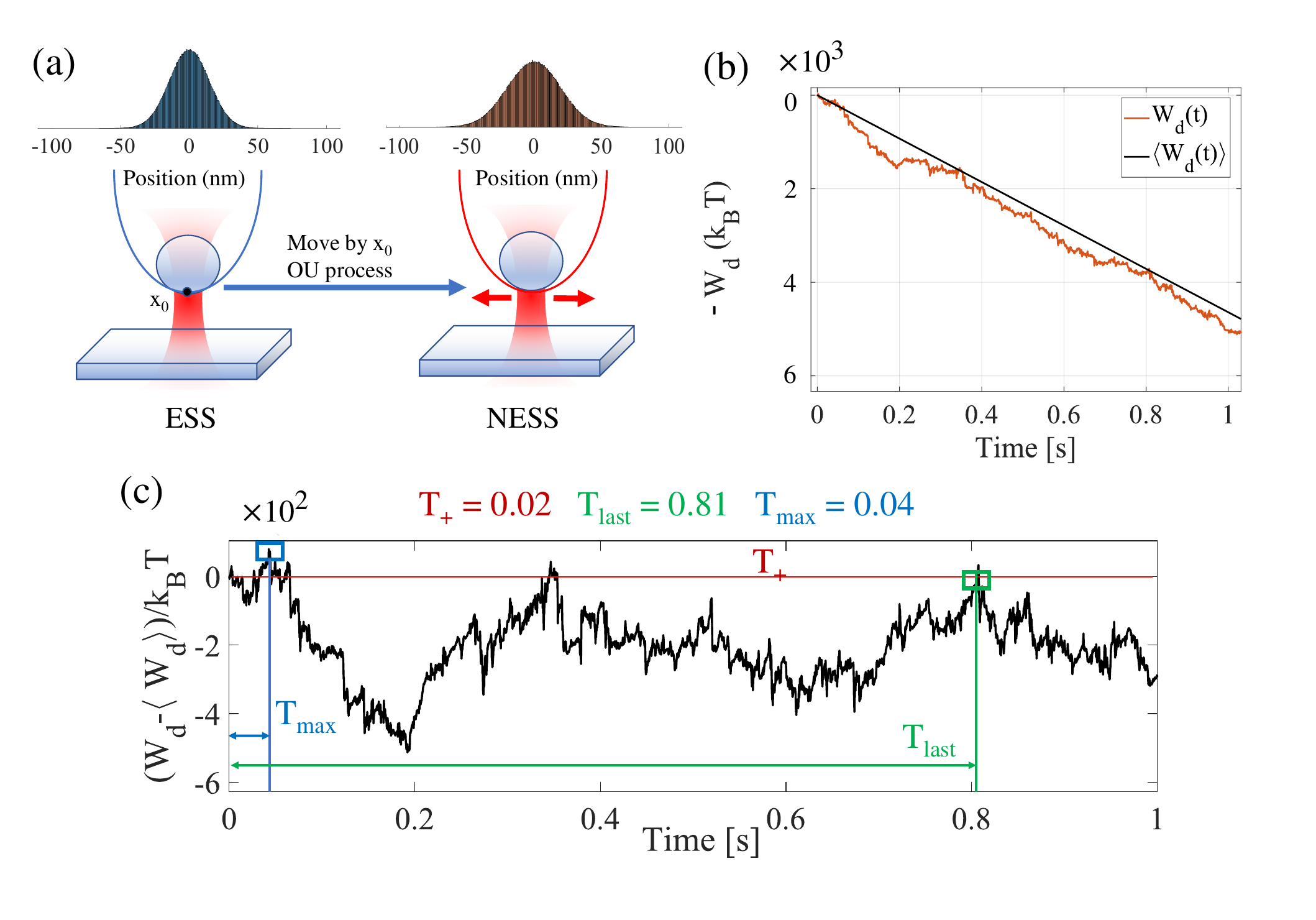}% Here is how to import EPS art
\caption{\label{fig:work}\textbf{Work done on the system and work dissipated by the system represented schematically, as obtained from experimental data.}
(a)Schematics of the process, where the trap center is modulated by an Ornstein Uhlenbeck process of a constant amplitude to proceed from an equilibrium to a non-equilibrium steady state (ESS to NESS) (b) The fraction of time the stochastic current $W_d$ (in red) lies above the average $\langle W_d\rangle$ over many cycles. (c) The work dissipated by the probe over time plotted in units of $k_BT$ after subtracting the mean value of dissipating current. Values of the L\'evy variables $T_+$,$T_{last}$,$T_{max}$ are displayed for this window in red, green, blue color coding, respectively.}  
\end{figure*}

\begin{figure*}[!t]
\includegraphics[width=12cm]{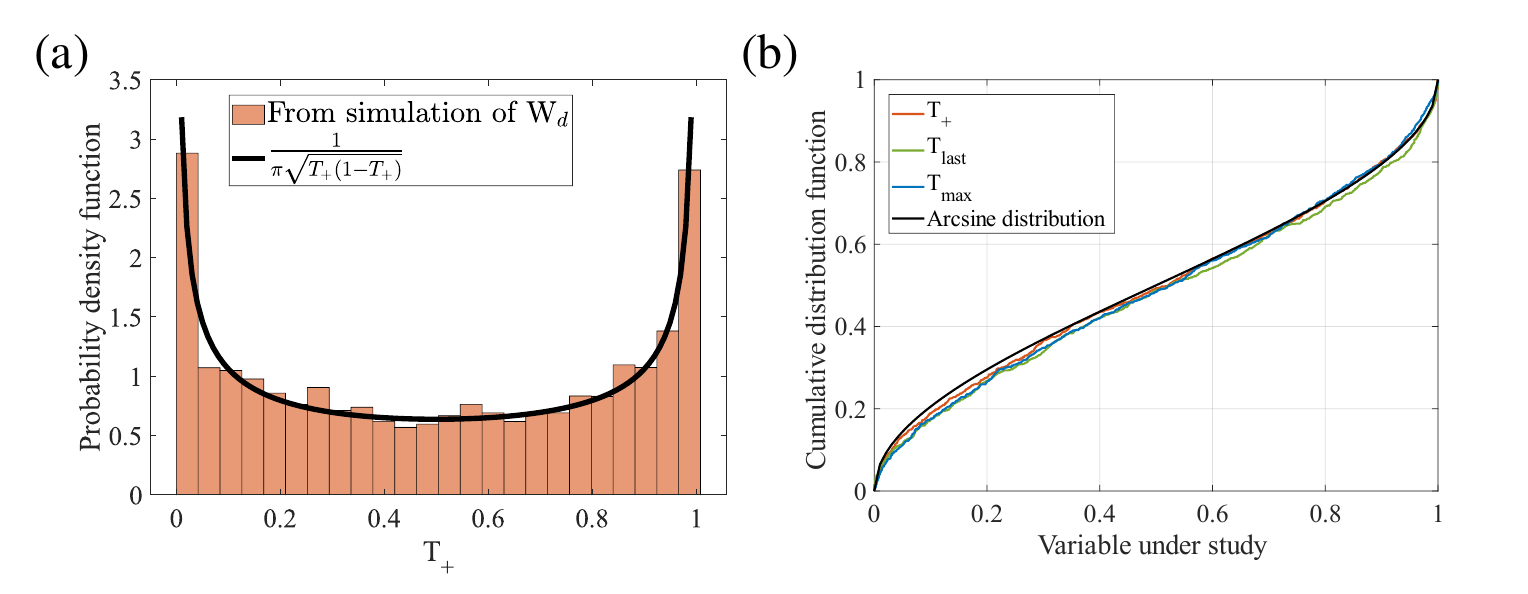}
\caption{\label{fig:simu} \textbf{Statistics of work dissipated as obtained from simulation.} The results for the three arcsine laws for $T_+$ (in red), $T_{last}$ (in green) and $T_{max}$ (in blue) from a simulation of the time series, where the CDF and PDF are plotted in \textbf{(a)} and \textbf{(b)}. [Parameters used for this example: $dt=1\ ms$, total time = 100 s, points in each division = 1000, $\tau_d=1.2\ ms$, $\tau_0=2.5\ ms$, $A=0.3\times(0.6\times10^{-6})^2\ m^2/s$  ]} 
\end{figure*}

As noted previously in \cite{barato2018arcsine}, fluctuations of any time integrated current, after a sufficiently large time $\tau$, obey an arcsine behavior, if the system is in a stationary state. Our model system, shown in Fig~\ref{fig:work}(a), is a probe modulated with Ornstein-Uhlenbeck noise, and forms a model NESS system that is expected to obey the properties of the first arcsine law. Now, in order to verify these laws, we first need to compute the stochastic currents (such as entropic current, work done or dissipated work) from the time series of the probe using Eq~\ref{eq:3} and Eq~\ref{eq:6}. In Fig~\ref{fig:work}(b) we show the work dissipated which we compute from a representative trajectory. In addition, we also demonstrate how the $T_+$, $T_{last}$ or $T_{max}$ variables are computed in Fig~\ref{fig:work}(c). We demonstrate the arcsine laws for this system using numerically simulated data. We adopt an Euler discretization scheme, with $\Delta t = 0.0001$, and generate a long steady state trajectory which is $1000s$ long. We obtain current statistics for any finite $\tau$ value by breaking up this long steady state trajectory into $N=1000/\tau$ number of trajectories of length $\tau$. We verify that the PDF and CDF of the variables computed on these trajectories follow $\frac{1}{\pi\sqrt{T(1-T)}}$ and the arcsine laws respectively, as we show in Fig~\ref{fig:simu} (a) and (b).\\ 

\begin{figure*}
    \centering
    \includegraphics[width=18 cm]{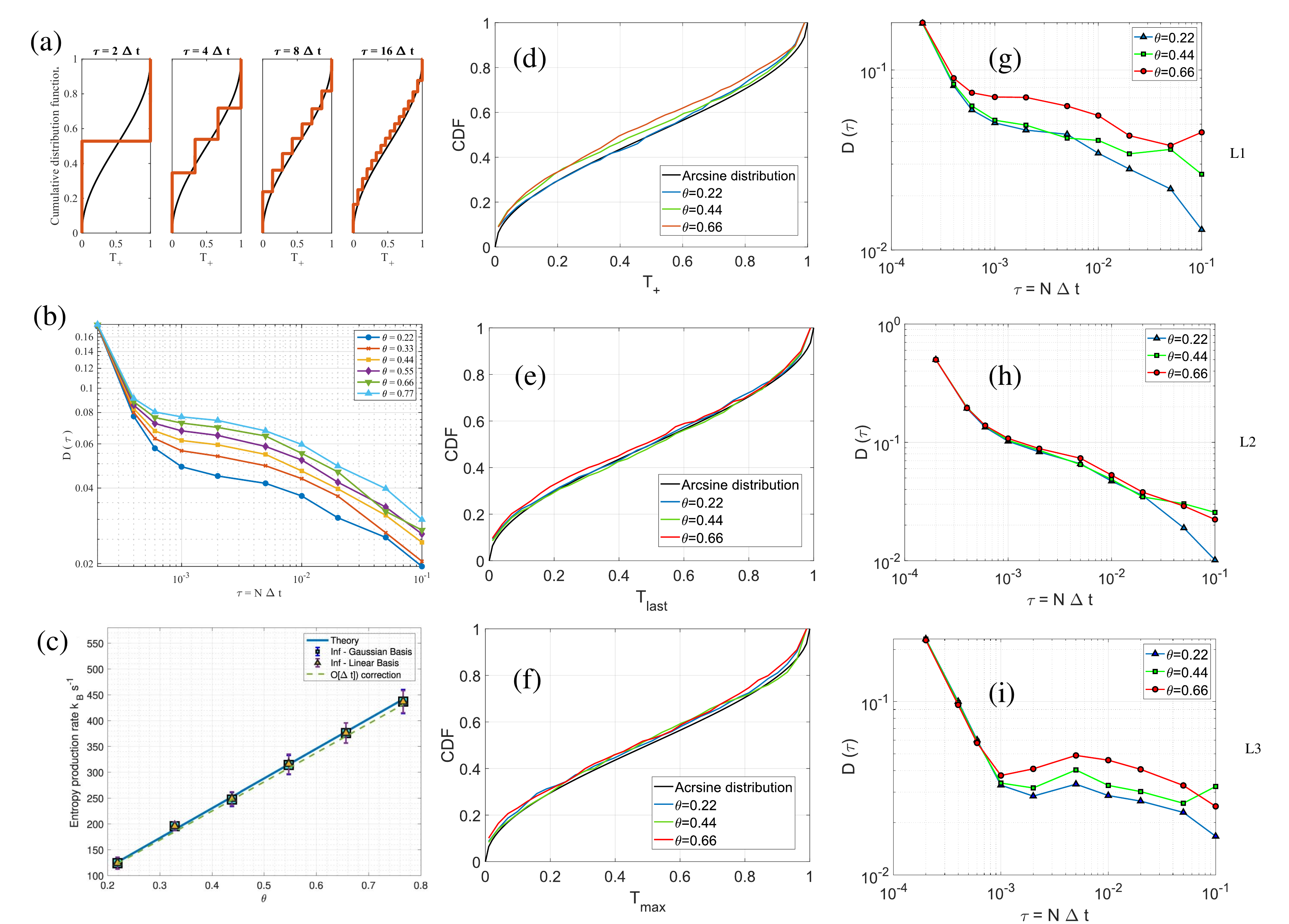}
\caption{\label{fig:ConvergencePlot} \textbf{Convergence to the arcsine law} (a) The cumulative distribution function of $T_+$ at finite $\tau$ values: at any finite $\tau = N\Delta t$, the probability distribution of $T_+$ is a discrete distribution due to the inherent discreteness of the non-equilibrium trajectory. The corresponding cumulative distributions look like step functions and converge to the smooth, arcsine law only asymptotically in $\tau$. (b) The $L_1$ measure of distance (Eq.\ \eqref{eq:distance}) between the instantaneous cumulative distribution and the asympotic arcsine law as a function of $\tau$, for different values of $\theta$. We find that, larger the value of $\theta$ (higher the entropy production rate), further the instantaneous cumulative distribution is from the arcsine law. (c) The rate of entropy production inferred increases with increasing  $\theta$ (reproduced from \cite{manikandan2021quantitative})
(d)-(f) The arcsine laws for $T_+$, $T_{last}$ and $T_{max}$ as observed from experimental trajectories of the stochastic sliding parabola model. (g)-(i) The $L_1$ measure of distance (Eq.\ \eqref{eq:distance}) between the instantaneous cumulative distribution and the asympotic arcsine law as a function of $\tau$, for different values of $\theta$.} 
\end{figure*}

\begin{figure*}
    \centering
    \includegraphics[width=16 cm]{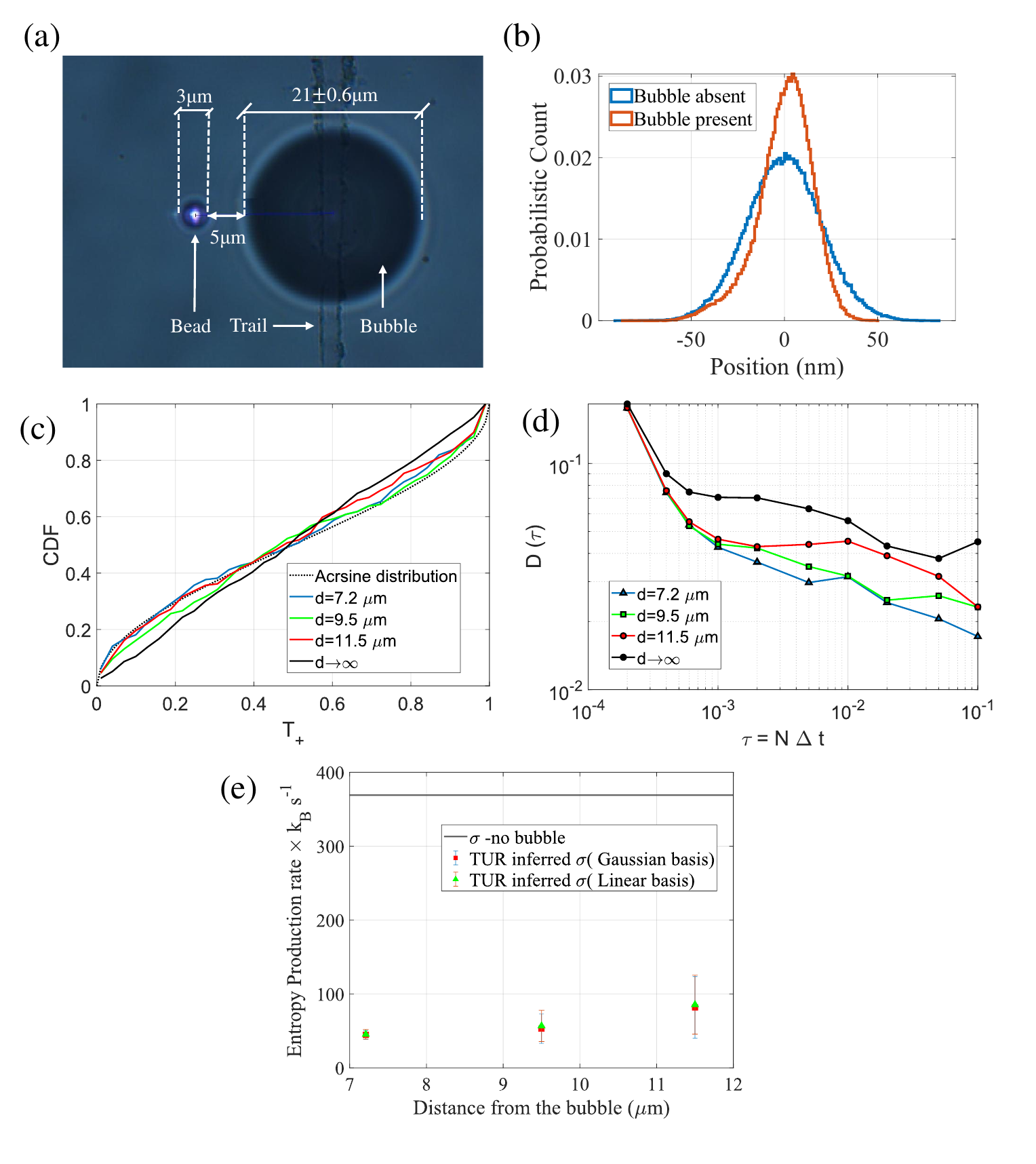}
\caption{\label{fig:bubble} \textbf{Action of a microbubble in close vicinity of the optically trapped probe.} {\textbf{(a)} Schematic of the experimental probe particle-micobubble system. The microbubble is grown on a pre-existing absorbing pattern/trail in the sample chamber (see Methods). The different dimensions of the bubble and }. \textbf{(b)} Histogram of the position fluctuation -- clearly it develops a skew in the presence of the flow field generated by the microbubble. \textbf{(c)} CDF of $T_+$ and \textbf{(d)} Convergence rates to the first arcsine law  for the colloidal probe placed at different distances from the microbubble. \textbf{(e)} Entropy production rate for changing distance of the microbubble. (see Ref~\cite{manikandan2021quantitative})  } 
\end{figure*}

Since the empirical trajectories of the system are discrete in practice, the admissible values of any of the variables $T_+$, $T_{last}$ or $T_{max}$ are discrete for any finite $\tau$. The corresponding probability density functions are therefore discrete distributions and their cumulative distributions appear like step functions. However, in the large $\tau$ limit, both the PDF as well as the CDF become smooth functions. In Fig.~\ref{fig:ConvergencePlot}a), we show this feature using the CDF of $T_+$ for different $\tau$ values. The convergence to the arcsine law, as a function of $\tau$ can be better assessed using distance functions $D(\tau)$ in the space of distributions \cite{lu2017nonequilibrium}. Here we use the $L_1$ measure of distance given by,
\begin{align}
\label{eq:distance}
    D(\tau) = \frac{1}{N_b}\sum_{i = 1}^{N_b} \vert \hat{P}^\tau_i -P_i\vert,
\end{align}
where $\hat{P}^\tau$ is the numerically estimated cumulative distribution function and $P$ is arcsine distribution. $N_b$ corresponds to the number of bins used along the $T$ axis. In Fig~\ref{fig:ConvergencePlot}b), from a simulated trajectory, we show how the distance function varies as a function of $\tau$ for different $\theta$ values. In all cases, the fluctuating current we consider is the corresponding $\Delta S_{tot}$. We see that larger the $\theta$ value, the further is the instantaneous cumulative distribution away from the arcsine law for any fixed $\tau$. From Eq.\ \eqref{eq:sigma}, we know that larger $\theta$ implies, higher entropy production rate and further away from equilibrium steady state. We conclude that closer the system is to equilibrium, \textit{faster} will be the convergence to the arcsine law.  \\ 
\subsection{Experiments}
Next, we proceed to check whether these features continue to exist in the experimental realization of the stochastic sliding parabola model. We use a Gaussian beam (1064 nm) which is tightly focused with a high numerical aperture of a standard oil immersion objective (100x, NA=1.3) an inverted microscope (Olympus IX71) to trap a spherical polystyrene particle (Sigma-Aldrich LB30, diameter = 3 $\mu$m), in a double-distilled aqueous solution, inside a custom sample chamber of thickness 100 $\mu$m, that we construct by inserting double-sided sticky tape between two coverslips, which we then mount on a motorized microscope stage. The laser passes through an acousto-optic modulator (BRIMROSE-AOM), and the first-order beam traps and modulates the particle perpendicular to the beam's direction, given by $U(x(t),\lambda(t))=k[x(t)-\lambda(t)]^2/2$,  where the trap stiffness is $k=19.7\pm0.1$ pN/$\mu$m. The input noise $\lambda(t)$ is generated through the Ornstein-Uhlenbeck process (see Eq~\ref{eq:lgv}) -- the time constant for its exponentially decaying correlation is $\tau_0=2.5$ ms, and its amplitude is chosen as $A= \left[0.1,0.2,0.3\right]\times(0.6\times10^{-6})^2\ m^2/s$. We focus a second co-propagating beam of wavelength 785 nm and measure the backscattered intensity by a balanced detection system constructed out of high gain-bandwidth photo-detectors (Thorlabs PDA100A2) \cite{bera2017fast} to sample the one-directional trajectory of the probe at a spatio-temporal resolution of 1 nm - 10 kHz for 100 seconds.\\

For the next stage of the experiment, we form a microbubble of diameter $21\pm0.6\ \mu$m (see Methods) using another laser \cite{ghosh2020directed} such that its surface remains $d=10\ \mu$m away from the mean position of the center of the probe particle.  We have already used this experimental setup to study one or more microbubbles with  colloidal particles moving in the liquid in different contexts \cite{ghosh2020directed,manikandan2021quantitative}. The microbubbles are nucleated on a liquid-glass interface. The surface is pre-coated by linear patterns of a Mb-based soft oxometalate (SOM) material \cite{roy2013langmuir}. When we focus a laser beam on any region along this pattern, the SOM material gets intensely heated and a microbubble forms. The top of the bubble is colder than its bottom where it is anchored to the interface. As the surface tension is a function of temperature, the variation of the surface tension along the surface of the bubble sets up a Marangoni stress, driving a flow along the surface of the bubble. The experimental apparatus is described in the Appendix. \cite{roy2013langmuir}.

\subsection{Discussions}
To obtain robust statistics, we divide entropic current computed from the 100 s position fluctuation time-series into segments of 1 s ($10^4$ points) each, which is much greater than the time constant of the underlying process, $\tau_d (=2.5\ ms)$. In Fig.~\ref{fig:ConvergencePlot}, we demonstrate convergence to the three arcsine laws as a function of $\tau$. The trends observed in Fig~\ref{fig:ConvergencePlot}(b) are reproduced quite well, in all the three cases of the arcsine laws as seen in Fig~\ref{fig:ConvergencePlot} (d)-(i). As seen in Ref~\cite{manikandan2021quantitative} and reproduced in Fig~\ref{fig:ConvergencePlot}(c), the rate of entropy production increases with increasing $\theta$, or the amplitude of the added noise. The convergence to the arcsine law is found to be faster for the near equilibrium system with the smallest entropy production rate, providing direct experimental evidence of the observations in numerical simulations. Following the argument by Barato \emph{et. al.}, in Ref.~\cite{barato2018arcsine}, we know that due to strong Markovian property, the increments of the stochastic current become independent random variables, for which the first arcsine law works in the first place. For the near equilibrium systems, when the strength of the added correlated noise is less, therefore the entropic current also has weaker temporal correlation, and should converge faster. We discuss this matter further in the Appendix, where we we describe how the inherent correlations in the noise can change the convergence rate of our system. The increased statistical fluctuations in the plots are due to having only $100s$ long trajectories for the statistical analysis as opposed to the $1000s$ long trajectories we used in the simulations.\\
\begin{figure*}
\includegraphics[width=0.99\textwidth]{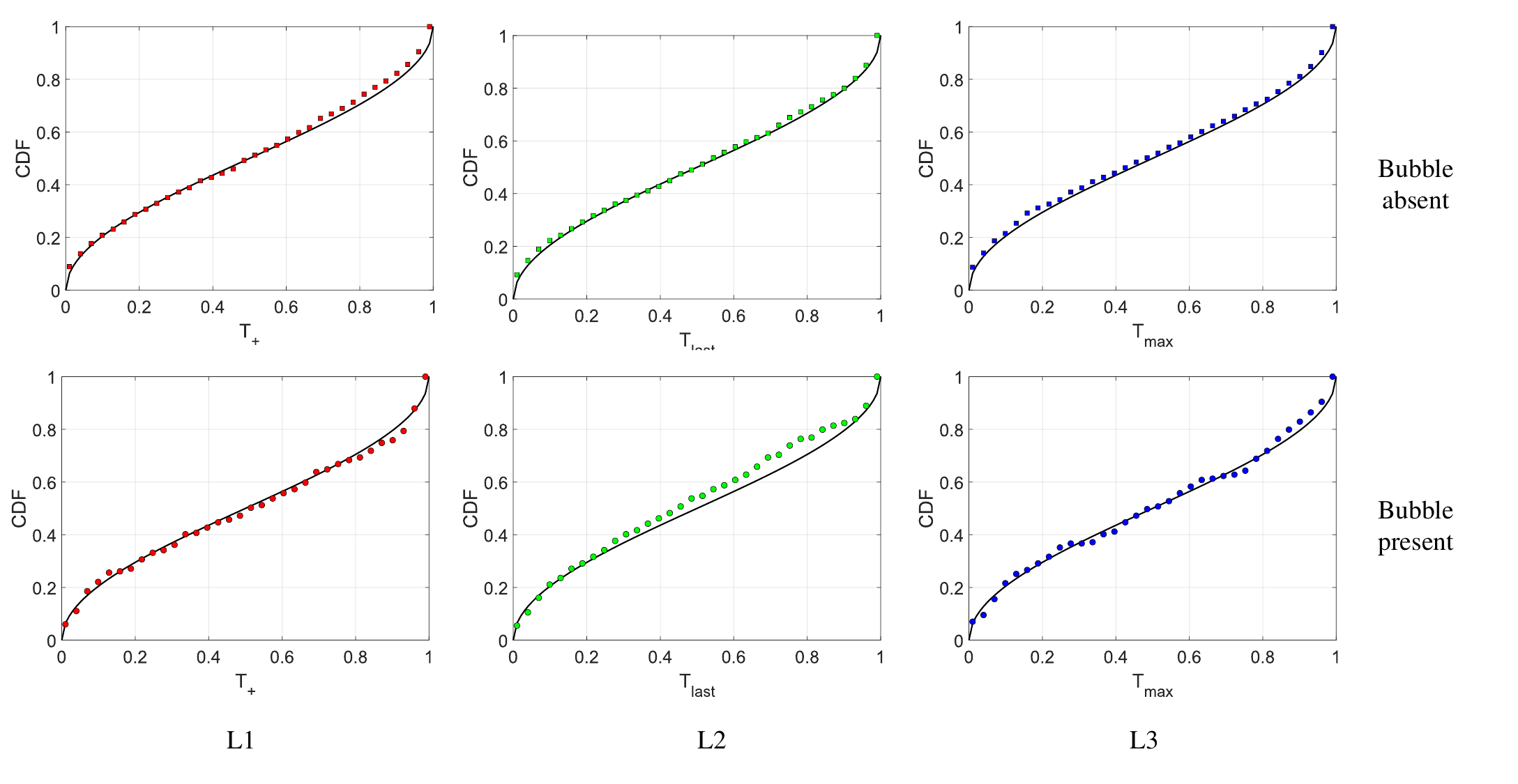}
\caption{\label{fig:asine} \textbf{Three arcsine laws tabulated for entropic currents.} The entropic currents computed from our experiments are used to plot the three arcsine laws for $T_+$, $T_{last}$ and $T_{max}$ in red, green and blue colours, respectively, for both the cases where bubble is present (in circles) and absent (in squares).}
\end{figure*}
For the second case, (Fig~\ref{fig:bubble}(a)) the fact that we perform the experiment in the proximity of the microscopic bubble, confines the trajectory colloidal particle and skews its probability distribution (see Fig~\ref{fig:bubble}(b)). The strength of the non-equilibrium currents is reduced in the presence of the micro-bubble, as explained in Ref~\cite{manikandan2021quantitative}. The decrease in entropy production rate decreases closer the colloidal system is to the bubble (see Fig~\ref{fig:bubble}(e)), and its trajectories look more reversible. We therefore consider the convergence to the (first) arcsine law as a function of $\tau$ with varying distances from the surface of the bubble. In Fig.~\ref{fig:bubble}(c-d), we find that the convergence to the arcsine law is faster when the colloidal setup is closer to the bubble, where the trajectories look more reversible. Here, we note that the thermodynamic force field $\textbf{F}(x)$ in the presence of the bubble is generically different from the one when there is no bubble. We nevertheless use the latter one for both cases, for simplicity (refer to \cite{manikandan2021quantitative} for details). In Fig.~\ref{fig:asine} we tabulate all the experimental plots and show that the CDF of $T_+$, $T_{last}$, and $T_{max}$ obtained from the entropic current follow L\'evy arcsine statistics. Additionally we note that change in the time-constant of the added gaussian noise also changes the rate of entropy production as given in Eq~\ref{eq:sigma}, and it similarly influences the convergence rate of the arcsine distribution. In the Appendix, we further show that the physical currents such as work done on the system or work dissipated by the system follow similar trends of the arcsine properties.

\subsection{Conclusions} We unambiguously demonstrate that a class of intrinsic stochastic currents associated with a Markov process, such as entropy, or physical currents like work done by, or dissipated from the system, in a non-equilibrium steady-state system obey L\'evy arcsine statistics. The cumulative probability distributions of the variables $T_+$, $T_{last}$ and $T_{last}$ converge to the arcsine distribution asymptotically in the long time limit and the rate of convergence is decided by the entropy generated by the stochastic currents. For the case where the amplitude of the colored noise \cite{saha2019stochastic,mestres2014realization} is low or an additional flow field is present due to a microbubble \cite{manikandan2021quantitative} at close proximity, the dissipated work is minimum and the rate of convergence is faster. Any deviations from arcsine laws indicate non-Markovian behavior \cite{sadhu2018generalized} or anomalous diffusion (``aging'') \cite{akimoto2020aging,weeks1998anomalous} in the underlying processes that generate the stochastic current. Thus it is possible to infer about underlying processes from the fluctuating currents, and thereby detect deviations from Gauss-Markov properties, or generalize it to include transient phenomena in physics, without scanning the system holistically. Understandably, this recipe may then be applied to problems such as diffusion on a crystal lattice, \cite{ala2002collective} kinesin walking on microtubules \cite{bugiel2018three} where studying long term statistics of stochastic currents is of utmost importance. Additionally, the arcsine laws are studied for long-time limits, where the time of observation far outruns the periodicity or correlation times of our given system. However, few attempts have been made to understand the statistics when the time period of observation is comparable to the inherent time scales of the system, i.e. before it has reached steady state. In that case, it is clear that the Markov properties are not satisfied and demands further investigation.

\section*{Acknowledgement}
We thank Dr. SK Manikandan, of NORDITA, KTH Royal Institute of Technology and Dr. S Krishnamurthy of Department of Physics, Stockholm University, Sweden, for their insightful discussions during various parts of the project and suggestions while preparing the manuscript. 

%\nocite{*}

%\bibliography{asine}% Produces the bibliography via BibTeX.

\begin{thebibliography}{35}%
\makeatletter
\providecommand \@ifxundefined [1]{%
 \@ifx{#1\undefined}
}%
\providecommand \@ifnum [1]{%
 \ifnum #1\expandafter \@firstoftwo
 \else \expandafter \@secondoftwo
 \fi
}%
\providecommand \@ifx [1]{%
 \ifx #1\expandafter \@firstoftwo
 \else \expandafter \@secondoftwo
 \fi
}%
\providecommand \natexlab [1]{#1}%
\providecommand \enquote  [1]{``#1''}%
\providecommand \bibnamefont  [1]{#1}%
\providecommand \bibfnamefont [1]{#1}%
\providecommand \citenamefont [1]{#1}%
\providecommand \href@noop [0]{\@secondoftwo}%
\providecommand \href [0]{\begingroup \@sanitize@url \@href}%
\providecommand \@href[1]{\@@startlink{#1}\@@href}%
\providecommand \@@href[1]{\endgroup#1\@@endlink}%
\providecommand \@sanitize@url [0]{\catcode `\\12\catcode `\$12\catcode
  `\&12\catcode `\#12\catcode `\^12\catcode `\_12\catcode `\%12\relax}%
\providecommand \@@startlink[1]{}%
\providecommand \@@endlink[0]{}%
\providecommand \url  [0]{\begingroup\@sanitize@url \@url }%
\providecommand \@url [1]{\endgroup\@href {#1}{\urlprefix }}%
\providecommand \urlprefix  [0]{URL }%
\providecommand \Eprint [0]{\href }%
\providecommand \doibase [0]{https://doi.org/}%
\providecommand \selectlanguage [0]{\@gobble}%
\providecommand \bibinfo  [0]{\@secondoftwo}%
\providecommand \bibfield  [0]{\@secondoftwo}%
\providecommand \translation [1]{[#1]}%
\providecommand \BibitemOpen [0]{}%
\providecommand \bibitemStop [0]{}%
\providecommand \bibitemNoStop [0]{.\EOS\space}%
\providecommand \EOS [0]{\spacefactor3000\relax}%
\providecommand \BibitemShut  [1]{\csname bibitem#1\endcsname}%
\let\auto@bib@innerbib\@empty
%</preamble>
\bibitem [{\citenamefont {Barato}\ \emph {et~al.}(2018)\citenamefont {Barato},
  \citenamefont {Rold{\'a}n}, \citenamefont {Mart{\'\i}nez},\ and\
  \citenamefont {Pigolotti}}]{barato2018arcsine}%
  \BibitemOpen
  \bibfield  {author} {\bibinfo {author} {\bibfnamefont {A.~C.}\ \bibnamefont
  {Barato}}, \bibinfo {author} {\bibfnamefont {{\'E}.}~\bibnamefont
  {Rold{\'a}n}}, \bibinfo {author} {\bibfnamefont {I.~A.}\ \bibnamefont
  {Mart{\'\i}nez}},\ and\ \bibinfo {author} {\bibfnamefont {S.}~\bibnamefont
  {Pigolotti}},\ }\href@noop {} {\bibfield  {journal} {\bibinfo  {journal}
  {Physical review letters}\ }\textbf {\bibinfo {volume} {121}},\ \bibinfo
  {pages} {090601} (\bibinfo {year} {2018})}\BibitemShut {NoStop}%
\bibitem [{\citenamefont {L{\'e}vy}(1940)}]{levy1940certains}%
  \BibitemOpen
  \bibfield  {author} {\bibinfo {author} {\bibfnamefont {P.}~\bibnamefont
  {L{\'e}vy}},\ }\href@noop {} {\bibfield  {journal} {\bibinfo  {journal}
  {Compositio mathematica}\ }\textbf {\bibinfo {volume} {7}},\ \bibinfo {pages}
  {283} (\bibinfo {year} {1940})}\BibitemShut {NoStop}%
\bibitem [{\citenamefont {Dale}\ and\ \citenamefont
  {Workman}(1980)}]{dale1980arc}%
  \BibitemOpen
  \bibfield  {author} {\bibinfo {author} {\bibfnamefont {C.}~\bibnamefont
  {Dale}}\ and\ \bibinfo {author} {\bibfnamefont {R.}~\bibnamefont {Workman}},\
  }\href@noop {} {\bibfield  {journal} {\bibinfo  {journal} {Financial analysts
  journal}\ }\textbf {\bibinfo {volume} {36}},\ \bibinfo {pages} {71} (\bibinfo
  {year} {1980})}\BibitemShut {NoStop}%
\bibitem [{\citenamefont {Baek}\ \emph {et~al.}(2008)\citenamefont {Baek},
  \citenamefont {Kim},\ and\ \citenamefont {Kim}}]{baek2008testing}%
  \BibitemOpen
  \bibfield  {author} {\bibinfo {author} {\bibfnamefont {S.~K.}\ \bibnamefont
  {Baek}}, \bibinfo {author} {\bibfnamefont {T.~Y.}\ \bibnamefont {Kim}},\ and\
  \bibinfo {author} {\bibfnamefont {B.~J.}\ \bibnamefont {Kim}},\ }\href@noop
  {} {\bibfield  {journal} {\bibinfo  {journal} {Physica A: Statistical
  Mechanics and its Applications}\ }\textbf {\bibinfo {volume} {387}},\
  \bibinfo {pages} {3660} (\bibinfo {year} {2008})}\BibitemShut {NoStop}%
\bibitem [{\citenamefont {Clauset}\ \emph {et~al.}(2015)\citenamefont
  {Clauset}, \citenamefont {Kogan},\ and\ \citenamefont
  {Redner}}]{clauset2015safe}%
  \BibitemOpen
  \bibfield  {author} {\bibinfo {author} {\bibfnamefont {A.}~\bibnamefont
  {Clauset}}, \bibinfo {author} {\bibfnamefont {M.}~\bibnamefont {Kogan}},\
  and\ \bibinfo {author} {\bibfnamefont {S.}~\bibnamefont {Redner}},\
  }\href@noop {} {\bibfield  {journal} {\bibinfo  {journal} {Physical Review
  E}\ }\textbf {\bibinfo {volume} {91}},\ \bibinfo {pages} {062815} (\bibinfo
  {year} {2015})}\BibitemShut {NoStop}%
\bibitem [{\citenamefont {Saigo}(2018)}]{saigo2018quantum}%
  \BibitemOpen
  \bibfield  {author} {\bibinfo {author} {\bibfnamefont {H.}~\bibnamefont
  {Saigo}},\ }in\ \href@noop {} {\emph {\bibinfo {booktitle} {Progress in
  Nanophotonics 5}}}\ (\bibinfo  {publisher} {Springer},\ \bibinfo {year}
  {2018})\ pp.\ \bibinfo {pages} {79--106}\BibitemShut {NoStop}%
\bibitem [{\citenamefont {S{\'a}nchez}\ \emph {et~al.}(2013)\citenamefont
  {S{\'a}nchez}, \citenamefont {Sothmann}, \citenamefont {Jordan},\ and\
  \citenamefont {B{\"u}ttiker}}]{sanchez2013correlations}%
  \BibitemOpen
  \bibfield  {author} {\bibinfo {author} {\bibfnamefont {R.}~\bibnamefont
  {S{\'a}nchez}}, \bibinfo {author} {\bibfnamefont {B.}~\bibnamefont
  {Sothmann}}, \bibinfo {author} {\bibfnamefont {A.~N.}\ \bibnamefont
  {Jordan}},\ and\ \bibinfo {author} {\bibfnamefont {M.}~\bibnamefont
  {B{\"u}ttiker}},\ }\href@noop {} {\bibfield  {journal} {\bibinfo  {journal}
  {New Journal of Physics}\ }\textbf {\bibinfo {volume} {15}},\ \bibinfo
  {pages} {125001} (\bibinfo {year} {2013})}\BibitemShut {NoStop}%
\bibitem [{\citenamefont {Schmiedl}\ and\ \citenamefont
  {Seifert}(2008)}]{schmiedl2008efficiency}%
  \BibitemOpen
  \bibfield  {author} {\bibinfo {author} {\bibfnamefont {T.}~\bibnamefont
  {Schmiedl}}\ and\ \bibinfo {author} {\bibfnamefont {U.}~\bibnamefont
  {Seifert}},\ }\href@noop {} {\bibfield  {journal} {\bibinfo  {journal} {EPL
  (Europhysics Letters)}\ }\textbf {\bibinfo {volume} {83}},\ \bibinfo {pages}
  {30005} (\bibinfo {year} {2008})}\BibitemShut {NoStop}%
\bibitem [{\citenamefont {Pigolotti}\ \emph {et~al.}(2017)\citenamefont
  {Pigolotti}, \citenamefont {Neri}, \citenamefont {Rold{\'a}n},\ and\
  \citenamefont {J{\"u}licher}}]{pigolotti2017generic}%
  \BibitemOpen
  \bibfield  {author} {\bibinfo {author} {\bibfnamefont {S.}~\bibnamefont
  {Pigolotti}}, \bibinfo {author} {\bibfnamefont {I.}~\bibnamefont {Neri}},
  \bibinfo {author} {\bibfnamefont {{\'E}.}~\bibnamefont {Rold{\'a}n}},\ and\
  \bibinfo {author} {\bibfnamefont {F.}~\bibnamefont {J{\"u}licher}},\
  }\href@noop {} {\bibfield  {journal} {\bibinfo  {journal} {Physical review
  letters}\ }\textbf {\bibinfo {volume} {119}},\ \bibinfo {pages} {140604}
  (\bibinfo {year} {2017})}\BibitemShut {NoStop}%
\bibitem [{\citenamefont {Bel}\ and\ \citenamefont
  {Barkai}(2006)}]{bel2006random}%
  \BibitemOpen
  \bibfield  {author} {\bibinfo {author} {\bibfnamefont {G.}~\bibnamefont
  {Bel}}\ and\ \bibinfo {author} {\bibfnamefont {E.}~\bibnamefont {Barkai}},\
  }\href@noop {} {\bibfield  {journal} {\bibinfo  {journal} {Physical Review
  E}\ }\textbf {\bibinfo {volume} {73}},\ \bibinfo {pages} {016125} (\bibinfo
  {year} {2006})}\BibitemShut {NoStop}%
\bibitem [{\citenamefont {Sadhu}\ \emph {et~al.}(2018)\citenamefont {Sadhu},
  \citenamefont {Delorme},\ and\ \citenamefont {Wiese}}]{sadhu2018generalized}%
  \BibitemOpen
  \bibfield  {author} {\bibinfo {author} {\bibfnamefont {T.}~\bibnamefont
  {Sadhu}}, \bibinfo {author} {\bibfnamefont {M.}~\bibnamefont {Delorme}},\
  and\ \bibinfo {author} {\bibfnamefont {K.~J.}\ \bibnamefont {Wiese}},\
  }\href@noop {} {\bibfield  {journal} {\bibinfo  {journal} {Physical review
  letters}\ }\textbf {\bibinfo {volume} {120}},\ \bibinfo {pages} {040603}
  (\bibinfo {year} {2018})}\BibitemShut {NoStop}%
\bibitem [{\citenamefont {Singh}\ and\ \citenamefont
  {Kundu}(2019)}]{singh2019generalised}%
  \BibitemOpen
  \bibfield  {author} {\bibinfo {author} {\bibfnamefont {P.}~\bibnamefont
  {Singh}}\ and\ \bibinfo {author} {\bibfnamefont {A.}~\bibnamefont {Kundu}},\
  }\href@noop {} {\bibfield  {journal} {\bibinfo  {journal} {Journal of
  Statistical Mechanics: Theory and Experiment}\ }\textbf {\bibinfo {volume}
  {2019}},\ \bibinfo {pages} {083205} (\bibinfo {year} {2019})}\BibitemShut
  {NoStop}%
\bibitem [{\citenamefont {Akimoto}\ \emph {et~al.}(2020)\citenamefont
  {Akimoto}, \citenamefont {Sera}, \citenamefont {Yamato},\ and\ \citenamefont
  {Yano}}]{akimoto2020aging}%
  \BibitemOpen
  \bibfield  {author} {\bibinfo {author} {\bibfnamefont {T.}~\bibnamefont
  {Akimoto}}, \bibinfo {author} {\bibfnamefont {T.}~\bibnamefont {Sera}},
  \bibinfo {author} {\bibfnamefont {K.}~\bibnamefont {Yamato}},\ and\ \bibinfo
  {author} {\bibfnamefont {K.}~\bibnamefont {Yano}},\ }\href@noop {} {\bibfield
   {journal} {\bibinfo  {journal} {Physical Review E}\ }\textbf {\bibinfo
  {volume} {102}},\ \bibinfo {pages} {032103} (\bibinfo {year}
  {2020})}\BibitemShut {NoStop}%
\bibitem [{\citenamefont {Mart{\'\i}nez}\ \emph {et~al.}(2016)\citenamefont
  {Mart{\'\i}nez}, \citenamefont {Rold{\'a}n}, \citenamefont {Dinis},
  \citenamefont {Petrov}, \citenamefont {Parrondo},\ and\ \citenamefont
  {Rica}}]{martinez2016brownian}%
  \BibitemOpen
  \bibfield  {author} {\bibinfo {author} {\bibfnamefont {I.~A.}\ \bibnamefont
  {Mart{\'\i}nez}}, \bibinfo {author} {\bibfnamefont {{\'E}.}~\bibnamefont
  {Rold{\'a}n}}, \bibinfo {author} {\bibfnamefont {L.}~\bibnamefont {Dinis}},
  \bibinfo {author} {\bibfnamefont {D.}~\bibnamefont {Petrov}}, \bibinfo
  {author} {\bibfnamefont {J.~M.}\ \bibnamefont {Parrondo}},\ and\ \bibinfo
  {author} {\bibfnamefont {R.~A.}\ \bibnamefont {Rica}},\ }\href@noop {}
  {\bibfield  {journal} {\bibinfo  {journal} {Nature physics}\ }\textbf
  {\bibinfo {volume} {12}},\ \bibinfo {pages} {67} (\bibinfo {year}
  {2016})}\BibitemShut {NoStop}%
\bibitem [{\citenamefont {Seifert}(2012)}]{seifert2012stochastic}%
  \BibitemOpen
  \bibfield  {author} {\bibinfo {author} {\bibfnamefont {U.}~\bibnamefont
  {Seifert}},\ }\href@noop {} {\bibfield  {journal} {\bibinfo  {journal}
  {Reports on progress in physics}\ }\textbf {\bibinfo {volume} {75}},\
  \bibinfo {pages} {126001} (\bibinfo {year} {2012})}\BibitemShut {NoStop}%
\bibitem [{\citenamefont {Bustamante}\ \emph {et~al.}(2005)\citenamefont
  {Bustamante}, \citenamefont {Liphardt},\ and\ \citenamefont
  {Ritort}}]{bustamante2005nonequilibrium}%
  \BibitemOpen
  \bibfield  {author} {\bibinfo {author} {\bibfnamefont {C.}~\bibnamefont
  {Bustamante}}, \bibinfo {author} {\bibfnamefont {J.}~\bibnamefont
  {Liphardt}},\ and\ \bibinfo {author} {\bibfnamefont {F.}~\bibnamefont
  {Ritort}},\ }\href@noop {} {\bibfield  {journal} {\bibinfo  {journal} {arXiv
  preprint cond-mat/0511629}\ } (\bibinfo {year} {2005})}\BibitemShut {NoStop}%
\bibitem [{\citenamefont {Grima}(2010)}]{grima2010effective}%
  \BibitemOpen
  \bibfield  {author} {\bibinfo {author} {\bibfnamefont {R.}~\bibnamefont
  {Grima}},\ }\href@noop {} {\bibfield  {journal} {\bibinfo  {journal} {The
  Journal of chemical physics}\ }\textbf {\bibinfo {volume} {133}},\ \bibinfo
  {pages} {07B604} (\bibinfo {year} {2010})}\BibitemShut {NoStop}%
\bibitem [{\citenamefont {Foglino}\ \emph {et~al.}(2019)\citenamefont
  {Foglino}, \citenamefont {Locatelli}, \citenamefont {Brackley}, \citenamefont
  {Michieletto}, \citenamefont {Likos},\ and\ \citenamefont
  {Marenduzzo}}]{foglino2019non}%
  \BibitemOpen
  \bibfield  {author} {\bibinfo {author} {\bibfnamefont {M.}~\bibnamefont
  {Foglino}}, \bibinfo {author} {\bibfnamefont {E.}~\bibnamefont {Locatelli}},
  \bibinfo {author} {\bibfnamefont {C.}~\bibnamefont {Brackley}}, \bibinfo
  {author} {\bibfnamefont {D.}~\bibnamefont {Michieletto}}, \bibinfo {author}
  {\bibfnamefont {C.}~\bibnamefont {Likos}},\ and\ \bibinfo {author}
  {\bibfnamefont {D.}~\bibnamefont {Marenduzzo}},\ }\href@noop {} {\bibfield
  {journal} {\bibinfo  {journal} {Soft matter}\ }\textbf {\bibinfo {volume}
  {15}},\ \bibinfo {pages} {5995} (\bibinfo {year} {2019})}\BibitemShut
  {NoStop}%
\bibitem [{\citenamefont {Gomez-Solano}\ \emph {et~al.}(2010)\citenamefont
  {Gomez-Solano}, \citenamefont {Bellon}, \citenamefont {Petrosyan},\ and\
  \citenamefont {Ciliberto}}]{gomez:ssw}%
  \BibitemOpen
  \bibfield  {author} {\bibinfo {author} {\bibfnamefont {J.~R.}\ \bibnamefont
  {Gomez-Solano}}, \bibinfo {author} {\bibfnamefont {L.}~\bibnamefont
  {Bellon}}, \bibinfo {author} {\bibfnamefont {A.}~\bibnamefont {Petrosyan}},\
  and\ \bibinfo {author} {\bibfnamefont {S.}~\bibnamefont {Ciliberto}},\
  }\href@noop {} {\bibfield  {journal} {\bibinfo  {journal} {EPL (Europhysics
  Letters)}\ }\textbf {\bibinfo {volume} {89}},\ \bibinfo {pages} {60003}
  (\bibinfo {year} {2010})}\BibitemShut {NoStop}%
\bibitem [{\citenamefont {Pal}\ and\ \citenamefont
  {Sabhapandit}(2013)}]{Pal:2013wfb}%
  \BibitemOpen
  \bibfield  {author} {\bibinfo {author} {\bibfnamefont {A.}~\bibnamefont
  {Pal}}\ and\ \bibinfo {author} {\bibfnamefont {S.}~\bibnamefont
  {Sabhapandit}},\ }\href {https://doi.org/10.1103/PhysRevE.87.022138}
  {\bibfield  {journal} {\bibinfo  {journal} {Phys. Rev. E}\ }\textbf {\bibinfo
  {volume} {87}},\ \bibinfo {pages} {022138} (\bibinfo {year}
  {2013})}\BibitemShut {NoStop}%
\bibitem [{\citenamefont {Verley}\ \emph {et~al.}(2014)\citenamefont {Verley},
  \citenamefont {Van~den Broeck},\ and\ \citenamefont
  {Esposito}}]{verley2014work}%
  \BibitemOpen
  \bibfield  {author} {\bibinfo {author} {\bibfnamefont {G.}~\bibnamefont
  {Verley}}, \bibinfo {author} {\bibfnamefont {C.}~\bibnamefont {Van~den
  Broeck}},\ and\ \bibinfo {author} {\bibfnamefont {M.}~\bibnamefont
  {Esposito}},\ }\href@noop {} {\bibfield  {journal} {\bibinfo  {journal} {New
  Journal of Physics}\ }\textbf {\bibinfo {volume} {16}},\ \bibinfo {pages}
  {095001} (\bibinfo {year} {2014})}\BibitemShut {NoStop}%
\bibitem [{\citenamefont {Manikandan}\ and\ \citenamefont
  {Krishnamurthy}(2017)}]{Manikandan:2017awd}%
  \BibitemOpen
  \bibfield  {author} {\bibinfo {author} {\bibfnamefont {S.~K.}\ \bibnamefont
  {Manikandan}}\ and\ \bibinfo {author} {\bibfnamefont {S.}~\bibnamefont
  {Krishnamurthy}},\ }\href {https://doi.org/10.1140/epjb/e2017-80432-9}
  {\bibfield  {journal} {\bibinfo  {journal} {The European Physical Journal B}\
  }\textbf {\bibinfo {volume} {90}},\ \bibinfo {pages} {258} (\bibinfo {year}
  {2017})}\BibitemShut {NoStop}%
\bibitem [{\citenamefont {Manikandan}\ and\ \citenamefont
  {Krishnamurthy}(2018)}]{manikandan2018exact}%
  \BibitemOpen
  \bibfield  {author} {\bibinfo {author} {\bibfnamefont {S.~K.}\ \bibnamefont
  {Manikandan}}\ and\ \bibinfo {author} {\bibfnamefont {S.}~\bibnamefont
  {Krishnamurthy}},\ }\href@noop {} {\bibfield  {journal} {\bibinfo  {journal}
  {Journal of Physics A: Mathematical and Theoretical}\ }\textbf {\bibinfo
  {volume} {51}},\ \bibinfo {pages} {11LT01} (\bibinfo {year}
  {2018})}\BibitemShut {NoStop}%
\bibitem [{\citenamefont {Manikandan}\ \emph {et~al.}(2021)\citenamefont
  {Manikandan}, \citenamefont {Ghosh}, \citenamefont {Kundu}, \citenamefont
  {Das}, \citenamefont {Agrawal}, \citenamefont {Mitra}, \citenamefont
  {Banerjee},\ and\ \citenamefont
  {Krishnamurthy}}]{manikandan2021quantitative}%
  \BibitemOpen
  \bibfield  {author} {\bibinfo {author} {\bibfnamefont {S.~K.}\ \bibnamefont
  {Manikandan}}, \bibinfo {author} {\bibfnamefont {S.}~\bibnamefont {Ghosh}},
  \bibinfo {author} {\bibfnamefont {A.}~\bibnamefont {Kundu}}, \bibinfo
  {author} {\bibfnamefont {B.}~\bibnamefont {Das}}, \bibinfo {author}
  {\bibfnamefont {V.}~\bibnamefont {Agrawal}}, \bibinfo {author} {\bibfnamefont
  {D.}~\bibnamefont {Mitra}}, \bibinfo {author} {\bibfnamefont
  {A.}~\bibnamefont {Banerjee}},\ and\ \bibinfo {author} {\bibfnamefont
  {S.}~\bibnamefont {Krishnamurthy}},\ }\href@noop {}  \ \Eprint
  {https://arxiv.org/abs/2102.11374} {arXiv:2102.11374 [cond-mat.soft]}(\bibinfo {year} {2021})
  \BibitemShut {NoStop}%
\bibitem [{\citenamefont {Secchi}\ \emph {et~al.}(2020)\citenamefont {Secchi},
  \citenamefont {Vitale}, \citenamefont {Mi{\~n}o}, \citenamefont {Kantsler},
  \citenamefont {Eberl}, \citenamefont {Rusconi},\ and\ \citenamefont
  {Stocker}}]{secchi2020effect}%
  \BibitemOpen
  \bibfield  {author} {\bibinfo {author} {\bibfnamefont {E.}~\bibnamefont
  {Secchi}}, \bibinfo {author} {\bibfnamefont {A.}~\bibnamefont {Vitale}},
  \bibinfo {author} {\bibfnamefont {G.~L.}\ \bibnamefont {Mi{\~n}o}}, \bibinfo
  {author} {\bibfnamefont {V.}~\bibnamefont {Kantsler}}, \bibinfo {author}
  {\bibfnamefont {L.}~\bibnamefont {Eberl}}, \bibinfo {author} {\bibfnamefont
  {R.}~\bibnamefont {Rusconi}},\ and\ \bibinfo {author} {\bibfnamefont
  {R.}~\bibnamefont {Stocker}},\ }\href@noop {} {\bibfield  {journal} {\bibinfo
   {journal} {Nature communications}\ }\textbf {\bibinfo {volume} {11}},\
  \bibinfo {pages} {1} (\bibinfo {year} {2020})}\BibitemShut {NoStop}%
\bibitem [{\citenamefont {Lu}\ and\ \citenamefont
  {Raz}(2017)}]{lu2017nonequilibrium}%
  \BibitemOpen
  \bibfield  {author} {\bibinfo {author} {\bibfnamefont {Z.}~\bibnamefont
  {Lu}}\ and\ \bibinfo {author} {\bibfnamefont {O.}~\bibnamefont {Raz}},\
  }\href@noop {} {\bibfield  {journal} {\bibinfo  {journal} {Proceedings of the
  National Academy of Sciences}\ }\textbf {\bibinfo {volume} {114}},\ \bibinfo
  {pages} {5083} (\bibinfo {year} {2017})}\BibitemShut {NoStop}%
\bibitem [{\citenamefont {Bera}\ \emph {et~al.}(2017)\citenamefont {Bera},
  \citenamefont {Paul}, \citenamefont {Singh}, \citenamefont {Ghosh},
  \citenamefont {Kundu}, \citenamefont {Banerjee},\ and\ \citenamefont
  {Adhikari}}]{bera2017fast}%
  \BibitemOpen
  \bibfield  {author} {\bibinfo {author} {\bibfnamefont {S.}~\bibnamefont
  {Bera}}, \bibinfo {author} {\bibfnamefont {S.}~\bibnamefont {Paul}}, \bibinfo
  {author} {\bibfnamefont {R.}~\bibnamefont {Singh}}, \bibinfo {author}
  {\bibfnamefont {D.}~\bibnamefont {Ghosh}}, \bibinfo {author} {\bibfnamefont
  {A.}~\bibnamefont {Kundu}}, \bibinfo {author} {\bibfnamefont
  {A.}~\bibnamefont {Banerjee}},\ and\ \bibinfo {author} {\bibfnamefont
  {R.}~\bibnamefont {Adhikari}},\ }\href@noop {} {\bibfield  {journal}
  {\bibinfo  {journal} {Scientific reports}\ }\textbf {\bibinfo {volume} {7}},\
  \bibinfo {pages} {1} (\bibinfo {year} {2017})}\BibitemShut {NoStop}%
\bibitem [{\citenamefont {Ghosh}\ \emph {et~al.}(2020)\citenamefont {Ghosh},
  \citenamefont {Ranjan}, \citenamefont {Das}, \citenamefont {Sen},
  \citenamefont {Roy}, \citenamefont {Roy},\ and\ \citenamefont
  {Banerjee}}]{ghosh2020directed}%
  \BibitemOpen
  \bibfield  {author} {\bibinfo {author} {\bibfnamefont {S.}~\bibnamefont
  {Ghosh}}, \bibinfo {author} {\bibfnamefont {A.~D.}\ \bibnamefont {Ranjan}},
  \bibinfo {author} {\bibfnamefont {S.}~\bibnamefont {Das}}, \bibinfo {author}
  {\bibfnamefont {R.}~\bibnamefont {Sen}}, \bibinfo {author} {\bibfnamefont
  {B.}~\bibnamefont {Roy}}, \bibinfo {author} {\bibfnamefont {S.}~\bibnamefont
  {Roy}},\ and\ \bibinfo {author} {\bibfnamefont {A.}~\bibnamefont
  {Banerjee}},\ }\href@noop {} {\bibfield  {journal} {\bibinfo  {journal} {Nano
  letters}\ }\textbf {\bibinfo {volume} {21}},\ \bibinfo {pages} {10} (\bibinfo
  {year} {2020})}\BibitemShut {NoStop}%
\bibitem [{\citenamefont {Roy}\ \emph {et~al.}(2013)\citenamefont {Roy},
  \citenamefont {Arya}, \citenamefont {Thomas}, \citenamefont {J{\"u}rgschat},
  \citenamefont {Venkata~Rao}, \citenamefont {Banerjee}, \citenamefont
  {Malla~Reddy},\ and\ \citenamefont {Roy}}]{roy2013langmuir}%
  \BibitemOpen
  \bibfield  {author} {\bibinfo {author} {\bibfnamefont {B.}~\bibnamefont
  {Roy}}, \bibinfo {author} {\bibfnamefont {M.}~\bibnamefont {Arya}}, \bibinfo
  {author} {\bibfnamefont {P.}~\bibnamefont {Thomas}}, \bibinfo {author}
  {\bibfnamefont {J.~K.}\ \bibnamefont {J{\"u}rgschat}}, \bibinfo {author}
  {\bibfnamefont {K.}~\bibnamefont {Venkata~Rao}}, \bibinfo {author}
  {\bibfnamefont {A.}~\bibnamefont {Banerjee}}, \bibinfo {author}
  {\bibfnamefont {C.}~\bibnamefont {Malla~Reddy}},\ and\ \bibinfo {author}
  {\bibfnamefont {S.}~\bibnamefont {Roy}},\ }\href
  {https://doi.org/10.1021/la402777e} {\bibfield  {journal} {\bibinfo
  {journal} {Langmuir}\ }\textbf {\bibinfo {volume} {29}},\ \bibinfo {pages}
  {14733} (\bibinfo {year} {2013})}\BibitemShut {NoStop}%
\bibitem [{\citenamefont {Saha}\ and\ \citenamefont
  {Marathe}(2019)}]{saha2019stochastic}%
  \BibitemOpen
  \bibfield  {author} {\bibinfo {author} {\bibfnamefont {A.}~\bibnamefont
  {Saha}}\ and\ \bibinfo {author} {\bibfnamefont {R.}~\bibnamefont {Marathe}},\
  }\href@noop {} {\bibfield  {journal} {\bibinfo  {journal} {Journal of
  Statistical Mechanics: Theory and Experiment}\ }\textbf {\bibinfo {volume}
  {2019}},\ \bibinfo {pages} {094012} (\bibinfo {year} {2019})}\BibitemShut
  {NoStop}%
\bibitem [{\citenamefont {Mestres}\ \emph {et~al.}(2014)\citenamefont
  {Mestres}, \citenamefont {Martinez}, \citenamefont {Ortiz-Ambriz},
  \citenamefont {Rica},\ and\ \citenamefont {Roldan}}]{mestres2014realization}%
  \BibitemOpen
  \bibfield  {author} {\bibinfo {author} {\bibfnamefont {P.}~\bibnamefont
  {Mestres}}, \bibinfo {author} {\bibfnamefont {I.~A.}\ \bibnamefont
  {Martinez}}, \bibinfo {author} {\bibfnamefont {A.}~\bibnamefont
  {Ortiz-Ambriz}}, \bibinfo {author} {\bibfnamefont {R.~A.}\ \bibnamefont
  {Rica}},\ and\ \bibinfo {author} {\bibfnamefont {E.}~\bibnamefont {Roldan}},\
  }\href@noop {} {\bibfield  {journal} {\bibinfo  {journal} {Physical Review
  E}\ }\textbf {\bibinfo {volume} {90}},\ \bibinfo {pages} {032116} (\bibinfo
  {year} {2014})}\BibitemShut {NoStop}%
\bibitem [{\citenamefont {Weeks}\ and\ \citenamefont
  {Swinney}(1998)}]{weeks1998anomalous}%
  \BibitemOpen
  \bibfield  {author} {\bibinfo {author} {\bibfnamefont {E.~R.}\ \bibnamefont
  {Weeks}}\ and\ \bibinfo {author} {\bibfnamefont {H.~L.}\ \bibnamefont
  {Swinney}},\ }\href@noop {} {\bibfield  {journal} {\bibinfo  {journal}
  {Physical Review E}\ }\textbf {\bibinfo {volume} {57}},\ \bibinfo {pages}
  {4915} (\bibinfo {year} {1998})}\BibitemShut {NoStop}%
\bibitem [{\citenamefont {Ala-Nissila}\ \emph {et~al.}(2002)\citenamefont
  {Ala-Nissila}, \citenamefont {Ferrando},\ and\ \citenamefont
  {Ying}}]{ala2002collective}%
  \BibitemOpen
  \bibfield  {author} {\bibinfo {author} {\bibfnamefont {T.}~\bibnamefont
  {Ala-Nissila}}, \bibinfo {author} {\bibfnamefont {R.}~\bibnamefont
  {Ferrando}},\ and\ \bibinfo {author} {\bibfnamefont {S.}~\bibnamefont
  {Ying}},\ }\href@noop {} {\bibfield  {journal} {\bibinfo  {journal} {Advances
  in Physics}\ }\textbf {\bibinfo {volume} {51}},\ \bibinfo {pages} {949}
  (\bibinfo {year} {2002})}\BibitemShut {NoStop}%
\bibitem [{\citenamefont {Bugiel}\ and\ \citenamefont
  {Sch{\"a}ffer}(2018)}]{bugiel2018three}%
  \BibitemOpen
  \bibfield  {author} {\bibinfo {author} {\bibfnamefont {M.}~\bibnamefont
  {Bugiel}}\ and\ \bibinfo {author} {\bibfnamefont {E.}~\bibnamefont
  {Sch{\"a}ffer}},\ }\href@noop {} {\bibfield  {journal} {\bibinfo  {journal}
  {Biophysical journal}\ }\textbf {\bibinfo {volume} {115}},\ \bibinfo {pages}
  {1993} (\bibinfo {year} {2018})}\BibitemShut {NoStop}%
\bibitem [{\citenamefont {Seifert}(2005)}]{PhysRevLett.95.040602}%
  \BibitemOpen
  \bibfield  {author} {\bibinfo {author} {\bibfnamefont {U.}~\bibnamefont
  {Seifert}},\ }\href {https://doi.org/10.1103/PhysRevLett.95.040602}
  {\bibfield  {journal} {\bibinfo  {journal} {Phys. Rev. Lett.}\ }\textbf
  {\bibinfo {volume} {95}},\ \bibinfo {pages} {040602} (\bibinfo {year}
  {2005})}\BibitemShut {NoStop}%
\end{thebibliography}
%\bibliographystyle{apsrev4-2}
\providecommand{\noopsort}[1]{}\providecommand{\singleletter}[1]{#1}%

\section{Appendix}
\setcounter{figure}{0}
A few aspects described briefly in the paper require additional clarification and experimental evidence for interested readers. In this Appendix, we comment on the effects of changes in the parameter of the processes, discuss the Wiener properties of our currents, comment on the addition of boundary terms, characterize the error in our measured value, and discuss the properties of the added noise.
\begin{figure*}
    \centering
    \includegraphics[width=13cm]{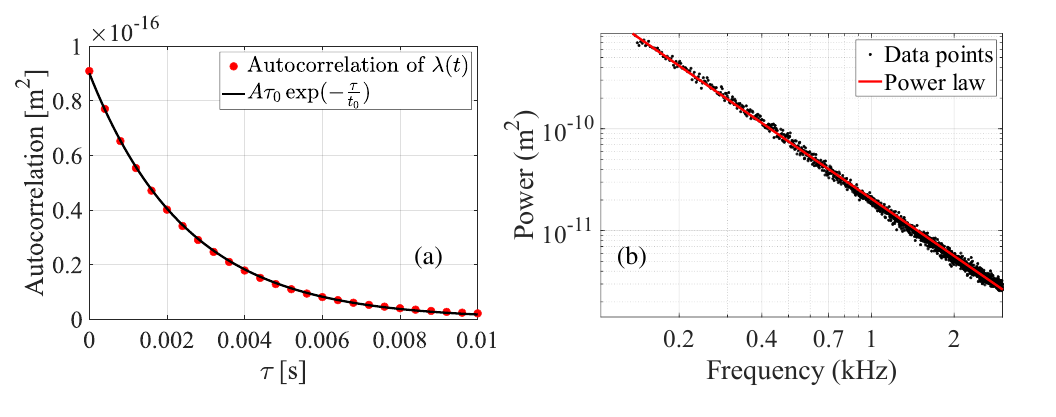}
    \caption{\textbf{Characteristics of the added Ornstein-Unhelbeck noise. (a)} The autocorrelation of the added noise which we fit with the theoretical exponential decay function with amplitude $A\tau_0$ and decay constant $\tau_0$. \textbf{(b)} The logarithmic plot of the power spectral density of the noise in the Fourier domain. The amplitude falls off as $1/f^\alpha$, where $\alpha$ is an exponent denoting the characteristics of the noise.}
    \label{fig:sup_lambda}
\end{figure*}

\subsection{Ornstein Uhlenbeck noise.}
Fig.~\ref{fig:sup_lambda} shows that the OU process generates a colored noise that has an exponentially decaying correlation and a power spectrum that fits a power law. We can use the autocorrelation property of the added noise to calibrate its amplitude. In Fig.~\ref{fig:sup_lambda}(a) we plot the autocorrelation for $\lambda(t)$ where $\tau_0=25\ ms$. In Fig.~\ref{fig:sup_lambda}(b), we plot the power spectrum of the noise in a logarithmic plot. By fitting the data points with $1/f^\alpha$, we get $\alpha\approx 1.8$, which confirms that it is colored noise.

\subsection{Markov property and convergence rates}
\begin{figure*}
    \centering
    \includegraphics[width=9cm]{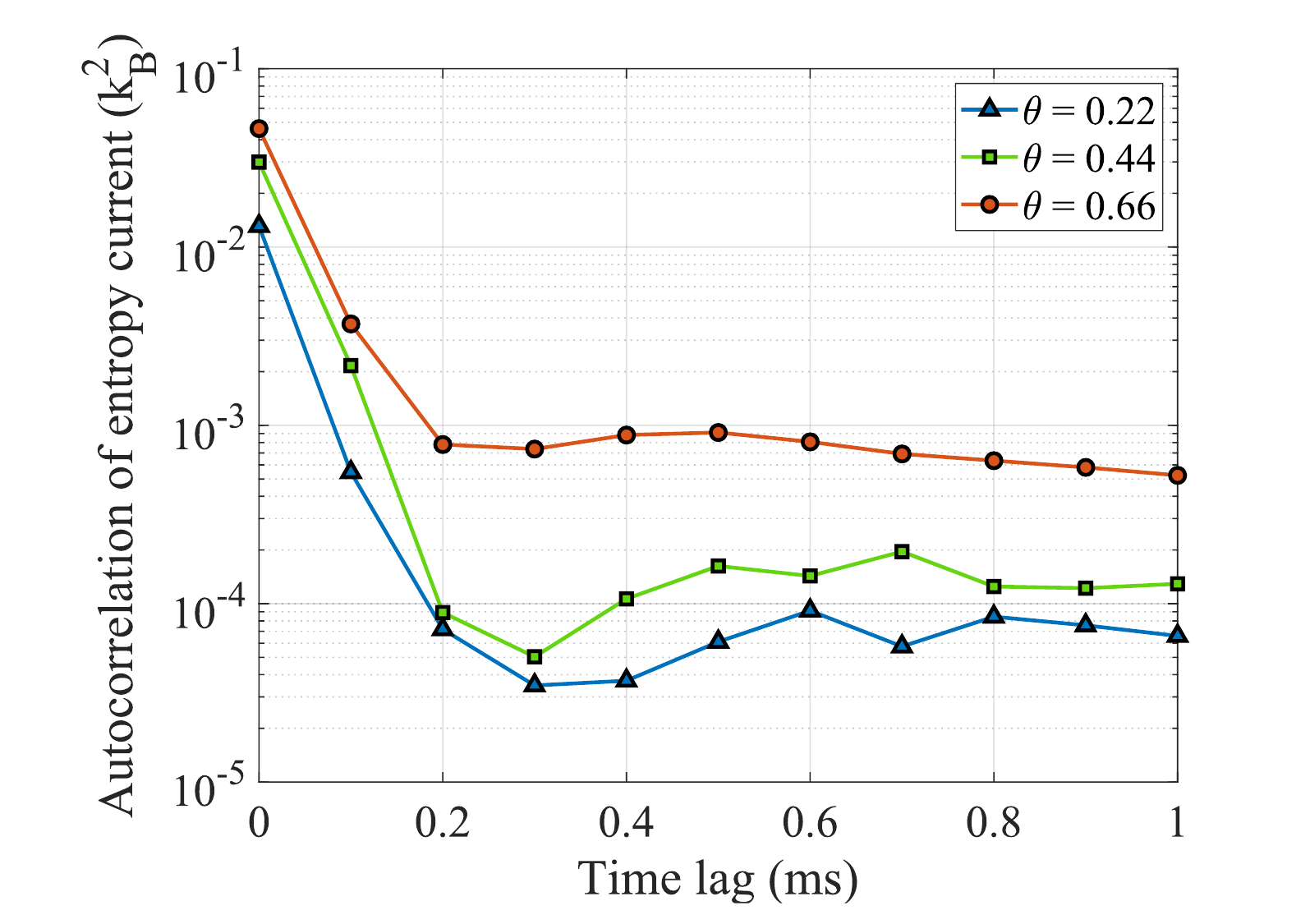}
    \caption{\textbf{Temporal autocorrelation of the entropy current} The temporal autocorrelation of the stochastic entropy current is plotted with respect to time delays. The correlation falls very fast with time-lag as it is a Markov process. For higher value of $\theta$ the time series for entropy is more strongly correlated.}
    \label{fig:autocorr_theta_vary}
\end{figure*}

\begin{figure*}[!t]
\includegraphics[width=15cm]{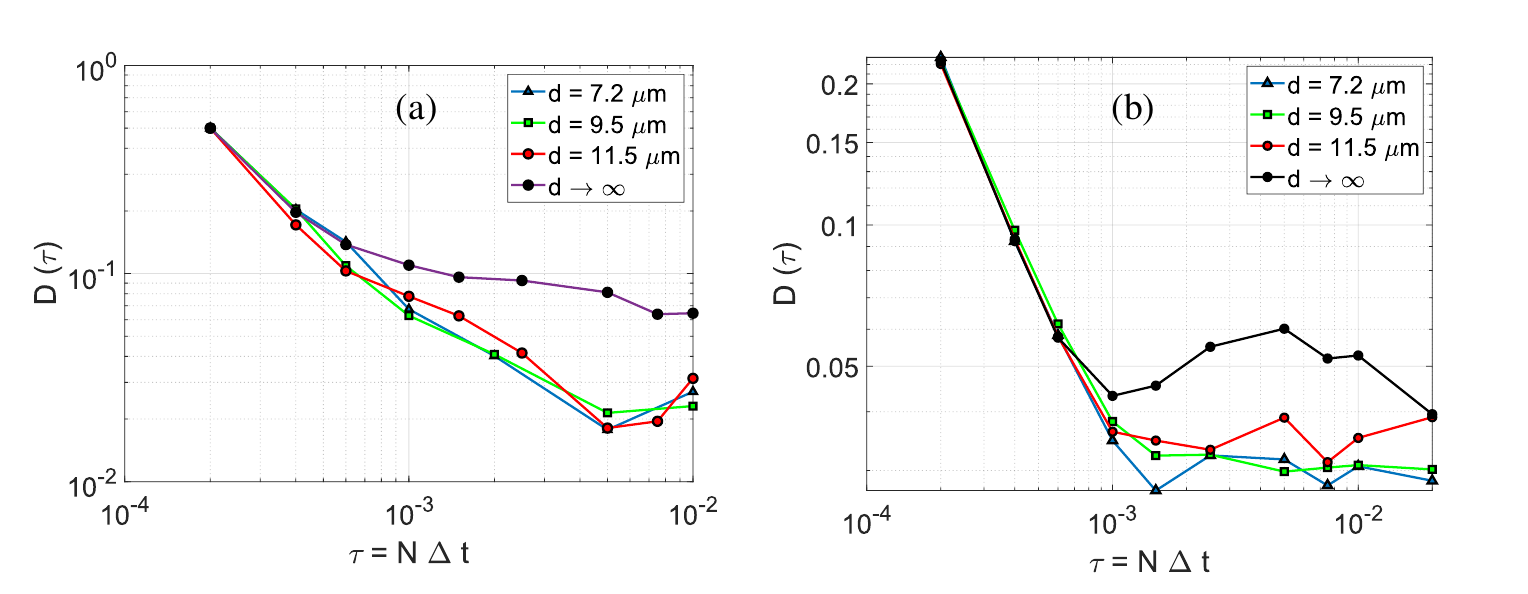}
\caption{\label{fig:buble_supple} \textbf{The second and third arcsine laws in presence of the bubble} The $L_1$ distance between instantaneous CDF and arcsine law is plotted as a function of $\tau$ for the $T_{last}$ and $T_{max}$, for various distances to the bubble and also for the case when there is no bubble present.}
\end{figure*}

In non-equilibrium setup, time integrated variables such as entropy produced or work done can be described by a discrete time master equation, with suitable transition rates between them. \cite{barato2018arcsine, PhysRevLett.95.040602}. The arcsine law is valid for these NESS variables because they can be successfully mapped to Markov chains that follow these acrsine properties. As we see in Fig.~\ref{fig:sup_lambda}(a), the added noise has non-Markov temporal correlation. Therefore the strength of the added noise contribute to the small non-Markov feature added to the trajectory of the stochastic current. Due to the strong Markov property we know that in the long time limit, the correlation would go to zero, hence the long time statistics are expected to follow the arcsine laws. Additionally as we confirm in our experiments and show in Fig~ \ref{fig:autocorr_theta_vary}, the trajectories with higher values of $\theta$ have stronger temporal correlation in their entropy current time series, and are expected to converge slower to the arcsine law. From Eq~\ref{eq:sigma}, we derive  $\sigma=\frac{\theta}{(1+\frac{\tau}{\tau_0})\tau}$, which shows that as the correlation strength of the added noise $\tau_0$ is increased, we would see a similar effect as increasing $\theta$, or amplitude of the added noise in our system. A similar effect is true for entropic currents produced by the probes at different distances to the bubble, which we have already shown in the main manuscript. The second and third arcsine laws are plotted for the same in Fig.~\ref{fig:buble_supple}
\subsection{Physical currents}
\begin{figure*}[!t]
    \includegraphics[width=15cm]{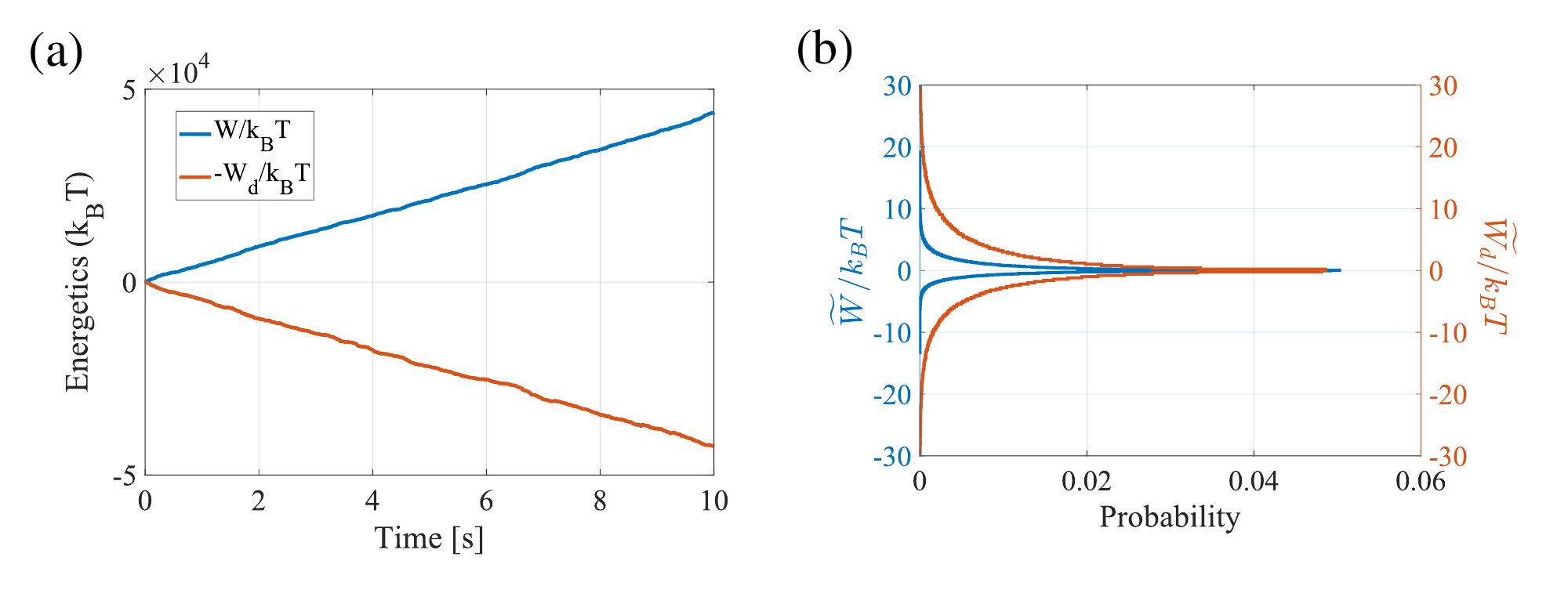}
    \caption{ The cumulative work done $(W)$ and work dissipated $(W_d)$ by the engine in units of $k_BT$ are plotted over time, which are positive and negative, respectively. \textbf{(a)} The probability distributions of the work done and work dissipated per unit sampling time $\widetilde{W}$ and $\widetilde{W_d}$ plotted in \textbf{(b)}}
    \label{fig:supple_wd}
\end{figure*}

\begin{figure*}[!t]
\includegraphics[width=15cm]{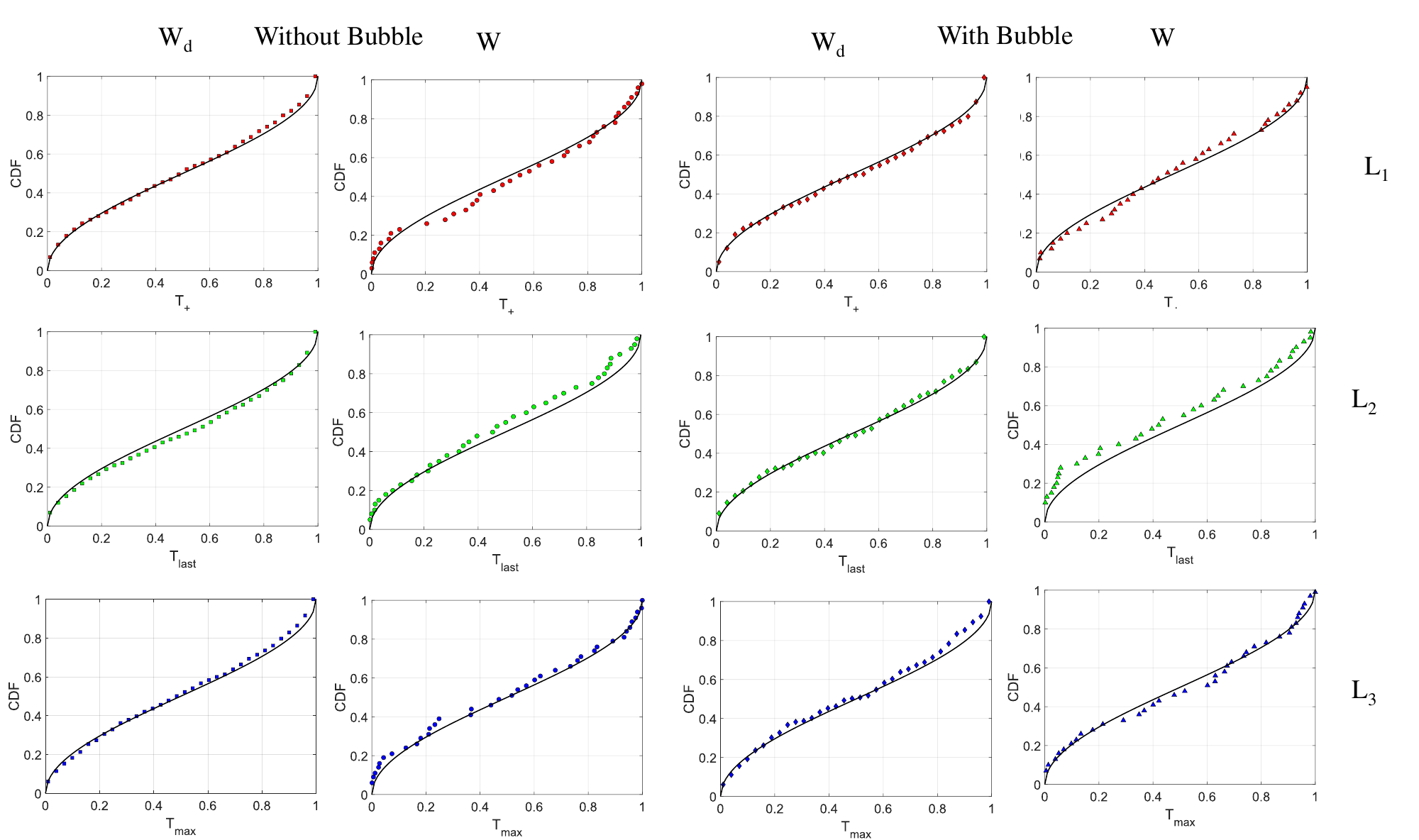}
\caption{\label{fig:work_asine} \textbf{arcsine laws tabulated for all the cases.} The work dissipated by the stochastic engine as well as the work done in our experiments are used to plot the three arcsine laws for $T_+$, $T_{last}$ and $T_{max}$ in red, green and blue colours, respectively.}
\end{figure*}

Any class of time-integrated current, from a NESS protocol will follow the arcsine laws, as long as their dynamics can be written in the form of Markov increments. Therefore physical currents in our system, such as the work done by the system as well as heat dissipated in the bath have quantifiable arcsine properties. Unlike the entropy production, which is an ensemble property of the trajectories, there "First-law" thermodynamic variables, are often more important to a wider class of experiments as they are more straightforward to calculate from the given dynamics. In Fig.~\ref{fig:supple_wd}(a) and (b), we show the CDF and PDF of the work done and heat dissipated by our colloidal system, respectively. Fig.~\ref{fig:work_asine} shows the arcsine properties of these variables. It is evident from the expression for work done that it stems from independent Markov increments - however, the boundary terms are needed to satisfy the Fluctuation dissipation theorem which can be calculated from the moment generating functions\cite{manikandan2018exact}\\

\begin{align*}
    W_d(\tau)&=\int_0^\tau \lambda(t)*dx(t)/dt\\&+ \frac{\delta^2(\delta(\theta(x_0^2-x^2_{\tau})+2x_0\lambda_0-2x_{\tau}\lambda_{\tau} -\lambda_0^2+\lambda_{\tau}^2)}{2D\tau_0(\delta^2(\theta+1)+2\delta+1)}
    \\&+\frac{2x_0\lambda_0-2x_{\tau}\lambda_{\tau}}{2D\tau_0(\delta^2(\theta+1)+2\delta+1)}
\end{align*}
We note that inclusion of higher order moments do not change the arcsine behaviour of the stochastic variable.

\subsection{Experimental details}
\textbf{Particle trapping and manipulation:} We perform all experiments on a double-distilled aqueous dispersion of spherical polystyrene particles (Sigma-Aldrich LB30) having a diameter 3 $\mu$m. We build a custom sample chamber of thickness 100 $\mu$m with double-sided sticky tape between two coverslips and mount the sample chamber on a motorized stage. A Gaussian beam (1064 nm) tightly focused with high numerical aperture oil immersion objective (100x, NA=1.3) of a standard oil immersion objective an inverted microscope (Olympus IX71) traps the particle at a height 15 $\mu$m from the lower surface of the sample chamber, to mitigate surface forces. The laser passes through an acousto-optic modulator (BRIMROSE-AOM), and the first-order beam traps and modulates the particle perpendicular to the beam's direction with input noise generated through the Ornstein-Uhlenbeck process. We focus a second co-propagating beam of wavelength 785 nm and measure the backscattered intensity by position-sensitive photo-detectors  (PDA100A2) employing a balanced detection system to sample the one-directional trajectory of the probe at a spatio-temporal resolution of 1 nm - 10 kHz.  Additionally, we use the autocorrelation of the time series of a trapped particle and the noise to calibrate the fluctuation of the probe from volts (measured by the photodiode) to nm.  

\textbf{Generation of microbubble:} For the experiments involving the microbubble, we employ a coverslip patterned by a polyoxometalate material \cite{ghosh2020directed}, absorbing at 1064 nm as the bottom surface of the sample chamber while keeping everything else unchanged. We use a second laser operating at a 1064 nm wavelength to generate a microbubble of diameter 21 $\mu$m -- the remains constant for unchanging laser power \cite{ghosh2020directed} throughout the time-scale of the experiment. We trap the particle at the same height as the radius of the bubble and calibrate the length scale of our setup with the company manufactured integrated software, which we also verify from the size of the probe.

\end{document}